\documentclass[conference,10pt]{IEEEtran}
\usepackage{graphicx,color}
\usepackage{amsmath, amsthm, amsfonts, amssymb, amsbsy,nccmath,bbm}
\usepackage{algorithm}
\usepackage{algpseudocode}
\usepackage{enumerate}
\usepackage{lipsum}
\usepackage{mathtools}

\usepackage[sort,compress]{cite}
\usepackage{epsfig}
\usepackage{epstopdf}
\usepackage{mathtools}
\usepackage{dsfont}
\usepackage{epstopdf}

\usepackage[inline]{enumitem}   
\makeatletter
\newcommand{\inlineitem}[1][]{%
\ifnum\enit@type=\tw@
    {\descriptionlabel{#1}}
  \hspace{\labelsep}%
\else
  \ifnum\enit@type=\z@
       \refstepcounter{\@listctr}\fi
    \quad\@itemlabel\hspace{\labelsep}%
\fi}
\makeatother
\newcommand{\beq}{\begin{equation}}
\newcommand{\eeq}{\end{equation}}

\def\adots{\mathinner{\mskip0mu\raise0pt\vbox{\kern7pt\hbox{.}}\mskip3mu
          \raise4pt\hbox{.}\mskip3mu\raise8pt\hbox{.}\mskip0mu}}

\newcommand{\bmV}{{\mathbf V}}

\newcommand{\tr}{\mbox{tr}}

\newcommand{\bmh}{\bfh}

\usepackage{bm}

\newcommand{\bme}{{\bm e}}
\newcommand{\bmx}{{\mathbf x}}
\newcommand{\bmy}{{\bm y}}
\newcommand{\bmv}{{\bm v}}

\newcommand{\bmH}{{\bm H}}

\newcommand{\bmR}{{\bm R}}

\renewcommand{\bmh}{{\bm h}}

\newcommand{\bmyt}{\widetilde{\bmy}}
\newcommand{\bmxt}{\widetilde{\bmx}}

\newcommand{\bmA}{{\bm A}}

\newcommand{\bmC}{{\bf C}}

\newcommand{\bmhh}{\widehat{\bmh}}
\newcommand{\bmI}{{\bf {I}}}
\newcommand{\bmD}{{\bf {D}}}
\newcommand{\bmY}{{\bf {Y}}}
\newcommand{\bmZ}{{\bf {Z}}}
\newcommand{\bmm}{{\bf {m}}}
\newcommand{\bmzero}{{\bm 0}}

\newcommand{\bmU}{\bm U}

\makeatletter
\newcommand\fs@spaceruled{\def\@fs@cfont{\bfseries}\let\@fs@capt\floatc@ruled
  \def\@fs@pre{\vspace{0.5\baselineskip}\hrule height.8pt depth0pt \kern2pt}%
  \def\@fs@post{\kern1pt\hrule\relax}%
  \def\@fs@mid{\kern2pt\hrule\kern2pt}%
  \let\@fs@iftopcapt\iftrue}
\makeatother

\newcommand{\bit}{\begin{itemize}}
\newcommand{\eit}{\end{itemize}}

\newcommand{\bmu}{\boldsymbol{\mu}}
\newcommand{\bmz}{\mathbf{z}}

\renewcommand{\bmh}{{\mathbf h}}
\renewcommand{\bmh}{{\mathbf h}}

\newcommand{\kb}{{\overline{k}}}

\newcommand{\vt}{{\widetilde{v}}}

\newcommand{\bmhb}{\overline{\bmh}}

\renewcommand{\bmA}{{\mathbf A}}

\newcommand{\bmX}{{\mathbf X}}
\newcommand{\bmht}{\widetilde{\bmh}}

\newcommand{\bmB}{{\mathbf B}}

\newcommand{\bmvt}{\widetilde{\bmv}}

\renewcommand{\bmU}{{\mathbf U}}

\newcommand{\bmE}{{\mathbf E}}

\newcommand{\bmr}{{\mathbf r}}

\newcommand{\bmvh}{{\widehat{\bmv}}}

\usepackage{lipsum}

\newcommand{\bmXi}{{\boldsymbol \Xi}}

\renewcommand{\bmzero}{{\boldsymbol 0}}

\newcommand{\bmdot}{{\boldsymbol .}}
\DeclareMathOperator{\E}{\mathbb{E}}

\renewcommand{\bmu}{{\bm u}}

\newcommand{\bmmu}{{\bm \mu}}

\newcommand{\pt}{{\widetilde{p}}}

\newcommand{\bmvb}{{\overline{\bmv}}}

\usepackage{mathabx}
\usepackage{amsmath}
\newcommand{\pc}{{\widecheck{p}}}
\newcommand{\bmhc}{{\widecheck{\bmh}}}

\newcommand{\xc}{{\widecheck{x}}}
\newcommand{\bmHb}{{\overline{\bmH}}}
\newcommand{\bmXb}{{\overline{\bmX}}}
\newcommand{\bmzb}{{\overline{\bmz}}}
\newcommand{\Lh}{{\widehat{L}}}

\newcommand{\xd}{{\dot{x}}}
\newcommand{\bmxd}{{\dot{\bmx}}}
\newcommand{\Kb}{{\overline{K}}}
\newcommand{\Gb}{{\overline{G}}}
\newcommand{\bmuh}{{\widehat{\bmu}}}

\newcommand{\bmUh}{{\widehat{\bmU}}}
\newcommand{\uh}{{\widehat{u}}}
\newcommand{\bmxb}{{\overline{\bmx}}}
\newcommand{\bmvc}{{\widecheck{\bmv}}}
\newcommand{\bmHc}{{\widecheck{\bmH}}}
\newcommand{\pbv}{{\breve{p}}}
\newcommand{\ptbv}{{\breve{\pt}}}

\newcommand{\bmhbv}{{\breve{\bmh}}}

\usepackage{subfiles}

\usepackage[acronym]{glossaries}

\newacronym{kld}{KLD}{Kullback–Leibler divergence}
\newacronym{snr}{SNR}{signal-to-noise ratio}
\newacronym{ap}{AP}{access point}

\begin{document}
\linespread{0.82}

\title{Distributed Iterative ML and Message Passing for Grant-Free Cell-Free Massive MIMO Systems}
\author{%
  \IEEEauthorblockN{Zilu Zhao$^*$, Christian Forsch$^\dagger$, Laura Cottatellucci$^\dagger$,  Dirk Slock$^*$}
  \IEEEauthorblockA{
			\small
			$^*$Communication Systems Department, EURECOM, Sophia Antipolis, France \\
            $^\dagger$Institute for Digital Communications, Friedrich-Alexander-Universit{\"a}t Erlangen-N{\"u}rnberg, Erlangen, Germany\\
			\{zilu.zhao, dirk.slock\}@eurecom.fr, \{christian.forsch, laura.cottatellucci\}@fau.de
			\vspace{-3mm}		}
	}

\maketitle

\begin{abstract}
Cell-Free (CF) Massive Multiple-Input Multiple-Output (MaMIMO) is considered one of the leading candidates for enabling next-generation wireless communication. With the growing interest in the Internet of Things (IoT), the Grant-Free (GF) access scheme has emerged as a promising solution to support massive device connectivity. The integration of GF and CF-MaMIMO introduces significant challenges, particularly in designing distributed algorithms for activity detection and pilot contamination mitigation.
In this paper, we propose a distributed algorithm that addresses these challenges. Our method first employs a component-wise iterative distributed Maximum Likelihood (ML) approach for activity detection, which considers both the pilot and data portions of the received signal. This is followed by a Pseudo-Prior Hybrid Variational Bayes and Expectation Propagation (PP-VB-EP) algorithm for joint data detection and channel estimation. Compared to conventional VB-EP, the proposed PP-VB-EP demonstrates improved convergence behavior and reduced sensitivity to initialization, especially when data symbols are drawn from a finite alphabet.
The pseudo prior used in PP-VB-EP acts as an approximated posterior and serves as a regularization term that prevents the Message Passing (MP) algorithm from diverging. To compute the pseudo prior in a distributed fashion, we further develop a distributed version of the Variable-Level Expectation Propagation (VL-EP) algorithm.

\end{abstract}
 
\vspace{-1mm}
\section{Introduction}
\vspace{-1mm}
Cell-Free (CF) Massive Multiple-Input Multiple-Output (MaMIMO) has attracted significant attention due to its ability to leverage the benefits of both MaMIMO and distributed antenna systems \cite{7827017, 9723203}. In CF MaMIMO systems, a large number of Access Points (APs) are geographically distributed over a wide area. Although a Central Processing Unit (CPU) coordinates the system, the information exchange between APs and the CPU is limited. Consequently, APs are unable to share their Channel State Information (CSI), which necessitates the development of distributed algorithms for data detection and channel estimation \cite{7827017, 10694527, karataev2024bilinear}.

With the sporadic traffic patterns introduced by the Internet of Things (IoT), the Grant-Free (GF) access scheme has been proposed as a reliable, low-latency solution \cite{8454392}. Moreover, it has been shown that MaMIMO systems are well-suited to support GF access.

Due to the large coverage area of CF MaMIMO systems and the absence of cellular boundaries, the number of User Terminals (UTs) often exceeds the number of available orthogonal pilot sequences. This results in the well-known issue of pilot contamination, which prevents the accurate acquisition of Channel State Information (CSI) \cite{9723203}. To address this challenge, semi-blind approaches—where Access Points (APs) jointly estimate the channel and detect user symbols—have been investigated in \cite{EURECOM+7308}. However, the joint estimation of CSI and data leads to an intractable high-dimensional inference problem, necessitating the use of approximate inference techniques.

\subsection{Prior Works}
In \cite{VBEP}, a hybrid Variational Bayes and Expectation Propagation (VB-EP) algorithm was proposed for semi-blind estimation by minimizing the Bethe Free Energy (BFE) \cite{9351786}. However, this method is sensitive to initialization and may diverge away from the stable point with bad initializations, particularly when user data are drawn from a finite alphabet. Moreover, it is not suitable for GF access schemes.

The Variable-Level EP (VL-EP) algorithm \cite{9723203} offers improved robustness when user data follow a Gaussian distribution. It can be interpreted as an approximate maximum a posteriori (MAP) estimator for the channel coefficients, due to its Expectation-Maximization (EM) update mechanism. However, VL-EP still suffers from divergence when finite alphabet data are used. Another limitation is that only a centralized version of this algorithm was proposed in \cite{9723203}. In contrast to VB-EP, VL-EP treats multiuser interference as zero-mean noise when estimating a user’s channel based on the data likelihood, which results in performance saturation in the high signal-to-noise ratio (SNR) regime.

To improve convergence, the concept of introducing a pseudo prior into message-passing (MP)-based algorithms has been proposed in \cite{pseudoprior}. A pseudo prior is essentially an approximate posterior, typically obtained from a simpler algorithm. In MP algorithms, it is incorporated by multiplying the pseudo prior with the true prior (or 1, if the parameter is deterministic) to form an enhanced prior. From this perspective, the pseudo prior does not introduce new information; rather, it serves as a regularization term that helps prevent the algorithm from diverging.

\subsection{Main Contributions}
In this paper, we propose a distributed algorithm for activity detection, channel estimation, and data detection. We observe that when pilot sequences are mutually orthogonal and the data symbols have zero mean, the Central Limit Theorem (CLT) can be applied to enable an approximate ML algorithm for distributed activity detection.
After the activity detection, we treat the activity indicator as given and perform the bilinear channel estimation and data detection jointly.
During the bilinear estimation, we leverage the concept of a pseudo prior to combine the strengths of VL-EP and VB-EP. We use the results in VL-EP as psedo prior for VB-EP to prevent VB-EP from drifting away from the critical point.

\label{Intro}
\vspace{-1mm}

\section{System Model}

We consider a semi-blind uplink signal model comprising $L$ APs and $\Kb$ single-antenna UTs. Each AP contains $N$ antennas. The local system model at the $l$-th AP is given by
\beq
\vspace{-2mm}
    \begin{bmatrix}
        \bmY_{p, l} &\bmY_{l}
    \end{bmatrix}
    =
    \bmHb_{l}\bmU
    \begin{bmatrix}
        \bmXb_{p} &\bmXb
    \end{bmatrix}
    +
    \begin{bmatrix}
        \bmV_{p, l} & \bmV_{l}
    \end{bmatrix}
    .
\eeq
The received signals at the the $l$-th AP consists of a pilot part $\bmY_{p, l} \in \mathbb{C}^{N \times P}$ and a data part $\bmY_{l}\in \mathbb{C}^{N \times T}$. Each column $\bmhb_{l\kb}$ of $\bmHb_{l}$ represents the uplink channel from UT $\kb$ to AP $l$. The channel $\bmhb_{l\kb}$ is modeled as a zero mean Gaussian random vector with covariance matrix $\bmXi_{\bmhb_{l\kb}}\in \mathbb{C}^{N\times N}$. 
The activity indicator matrix is denoted by $\bmU=\text{diag}[u_1, \dots, u_{\Kb}]$, which is an unknown deterministic diagonal matrix with binary entries.  Its diagonal elements form a vector $\bmu=\text{vec}(\bmU)$.
The transmitted signal matrix is composed of  pilot symbols $\bmXb_{p}\in \mathbb{C}^{\Kb \times P}$ and data symbols $\bmXb\in \mathcal{S}^{\Kb\times T}$, where $\mathcal{S}$ denotes the constellation set. 
We denote $\bmxb_{\kb}^T$ as the $\kb$-th row of $\bmXb$, and it represents the data sequence transmitted by user $\kb$.
The signal power is denoted as $\sigma_x^2$. The noise is considered as i.i.d. Gaussian distribution, and thus, $\text{vec}(\begin{bmatrix}
        \bmV_{p, l} & \bmV_{l}
    \end{bmatrix})\sim\mathcal{CN}(0, \bmC_{v})$ with $\bmC_{v}=\sigma_{v}^2\bmI$.
We denote $\bmY_p$, $\bmY$, $\bmHb$, $\bmV_p$ and $\bmV$ as global system paramenter.

\subsection {Orthogonal Pilot sequences}
If orthogonal pilot sequences are used, 
we can first preprocess the pilot observation by right multiplying it with $\bmxt_{p, g}^*$ which is the conjugated $g$-th pilot sequence from the pilot codebook. This results in an equivalent observation $\bmyt_{p, lg}$
\beq
    \bmyt_{p, lg}=\frac{1}{P\sigma_{x}^2}\bmY_{p, l}\bmxt_{p, g}^*=\sum_{\kb\in \Gb_g}\bmhb_{l\kb}u_{\kb}+\bmvt_{p, lg}
\eeq
where $\bmvt_{p, lg}=\frac{1}{P\sigma_{x}^2}\bmV_{p, l}\bmxt_{p, g}^*\sim \mathcal{N}(\bmv_{p, lg}|\bmzero, \bmC_{\bmvt})$, $\bmC_{\bmvt}=\frac{\sigma_v^2\bmI}{P\sigma_x^2}$, and $\Gb_g$ denotes the set of users using the $g$-th pilot sequence. We observe that every $\bmhb_{l\kb}u_{\kb}$ occurs only in one group $\Gb_g$, and the cross-correlation $\E[\bmvt_{p,lg}\bmvt_{p, lg'}^H]$ is an all-zero matrix for all $g\neq g'$. Therefore, the observations $\bmyt_{p, lg}$ and $\bmyt_{p, lg'}$ are independent. Denote $\bmHb_{l\Gb_g}$ as the matrix stacking $\forall \kb\in \Gb_g, \bmhb_{l\kb}$ as columns.

\section{Iterative ML for User Activity Detection}
The method we use for activity detection is based on maximum likelihood. 
Denote the estimated activity indicator as $\bmuh=\text{Diag}[\bmUh]$. The likelihood of $\bmuh$ is
\beq
\begin{split}
    p(\bmY, \bmY_p|\bmUh)=\int p(\bmY|\bmHb, \bmUh, \bmXb) p(\bmXb) \\
    \cdot\prod_{l} p(\bmHb_l) \prod_g p(\bmyt_{p, lg}|\bmHb_{l\Gb_g}, \bmUh) d\bmHb d\bmXb
\end{split}
\label{eq:wsa2506}
\eeq
Before investigating \eqref{eq:wsa2506}, we introduce a useful relation
\beq
\begin{split}
    &\mathcal{CN}(\bmm_1|\bmA\bmx, \bmC_1) \mathcal{CN}(\bmx|\bmm_2, \bmC_2)\\\
    &=\mathcal{CN}(\bmx|\bmm_3, \bmC_3)\mathcal{CN}(\bmm_1|\bmA\bmm_2, \bmC_1+\bmA\bmC_2\bmA^H)
\end{split}
\eeq
where 
\beq
    \begin{split}
        &\bmC_3=(\bmA^H\bmC_1^{-1}\bmA+\bmC_2^{-1})^{-1}\\
        &\bmm_3=\bmC_3(\bmA^H\bmC_1^{-1}\bmm_1+\bmC_2^{-1}\bmm_2).
    \end{split}
\eeq
This relation can be verified by noticing 
\beq
\begin{split}
\vspace{-1mm}
    \bmD(\bmA+\bmB\bmC\bmD)^{-1}=\bmC^{-1}(\bmD\bmA^{-1}\bmB+\bmC^{-1})^{-1}\bmD\bmA^{-1
    };\\
    \nonumber
    \det(\bmA+\bmB\bmC\bmD)=\det(\bmA)\det(\bmB)\det(\bmC^{-1}+\bmD\bmA^{-1}\bmB).
\end{split}
\eeq
We combine pilot observation and channel prior to get a pilot based prior $\pt(\bmHb_{l\Gb_g}|\bmuh)$ (even though it is a posterior given $\bmY_p$, we can still understand it as prior information as if we have a blind estimation problem)
\beq
\begin{split}
    \pt(\bmHb_{l\Gb_g}|\bmuh)=p(\bmHb_{l\Gb_g}|\bmY_p, \bmuh)\propto p(\bmyt_{p, lg}, \bmhb_{l\Gb_g}|\bmuh)\\
    = \mathcal{CN}(\bmyt_{p, lg}|\sum_{\kb\in \Gb_g\cap U(\bmuh)}\bmhb_{l\kb}, \bmC_{\vt})\mathcal{CN}(\bmhb_{l\Gb_g}|\bmzero, \bmXi_{\bmhb_{l\Gb_{g}}})
\end{split}
\eeq
Assume $\kb\in \Gb_g$, and in later context, we will use the marginal pilot-based prior $\pt(\bmhb_{l\kb})$ which is computed from
\vspace{-1mm}
\beq
\vspace{-5mm}
    \begin{split}
        \int p(\bmyt_{p, lg}, \bmhb_{l\Gb_g}|\bmuh) d\bmhb_{l\{\kb'\neq \kb\}}\!\!=\!\mathcal{CN}(\bmyt_{p, lg}|\bmzero, \bmC_{\bmyt_{p, lg}|\bmuh})\pt(\bmhb_{l\kb}|\bmuh)
    \end{split}
    \nonumber
\eeq
where
\vspace{-2mm}
\beq
\begin{split}
    &\bmC_{\bmyt_{p, lg}|\bmuh}=\bmC_{\vt}+\sum_{\kb\in \Gb_g\cap U(\bmuh)}\bmXi_{\bmhb_{l\kb}};\\
    &\pt(\bmhb_{l\kb}|\bmuh)=p(\bmhh_{l\kb}|\bmuh, \bmY_p)=\mathcal{CN}(\bmhb_{l\kb}|\bmm_{\bmhb_{l\kb}|\bmuh}, \bmC_{\bmhb_{l\kb}|\bmuh})
\end{split}
\eeq
and
\beq
    \begin{split}
        \bmC_{\bmY_p|\bmhb_{l\kb}, \bmuh}=\bmC_{\bmyt_{p, lg}|\bmuh}-\bmXi_{\bmhb_{l\kb}};\\
        \bmC_{\bmhb_{l\kb}|\bmuh}=(\bmC_{\bmY_p|\bmhb_{l\kb}, \bmuh}^{-1}+\bmXi_{\bmhb_{l\kb}}^{-1})^{-1}\\
        \bmm_{\bmhb_{l\kb}|\bmuh}=\bmC_{\bmhb_{l\kb}|\bmuh}\bmC_{\bmY_p|\bmhb_{l\kb}}^{-1}\bmyt_{p, lg}.
    \end{split}
    \label{eq:aksdfjgv}
\eeq

Now we take data observation into consideration. Rewrite \eqref{eq:wsa2506},
\beq
\begin{split}
     p(\bmY, \bmY_p|\bmUh)=\left[\prod_{l, g}\mathcal{CN}(\bmyt_{p, lg}|\bmzero, \bmC_{\bmyt_{p, lg}|\bmuh})\right]\\ 
     \cdot\int p(\bmY|\sum_{\kb\in U(\bmuh)}\bmxb_{\kb}\otimes\bmhb_{\kb}) p(\bmXb) 
    \cdot\prod_{l, g} \pt(\bmHb_{l\Gb_g}|\bmUh) d\bmH d\bmX
\end{split}
\eeq
where we use CLT to approximate the summation. Suppose that $\kb\in U(\bmuh)$, and we stack the results from \eqref{eq:aksdfjgv} to form
\beq
    \begin{split}
        &\bmC_{\bmhb_{\kb}|\bmuh}:=\text{cov}[\bmhb_{\kb}]\Rightarrow [\bmC_{\bmhb_{\kb}|\bmuh}]_{ll'}^{(N)}=\delta_{ll'}\bmC_{\bmhb_{l\kb}|\bmuh}\\
        &\bmm_{\bmhb_{\kb}|\bmuh}:=\E[\bmhb_{\kb}]
        \Rightarrow [\bmm_{\bmhb_{\kb}|\bmuh}]_{l}^{(N)}=\bmm_{\bmhb_{l\kb}|\bmuh}
    \end{split}
\eeq
The cross correlation between different $\bmzb_{\kb}$ is not important, since $\forall \kb\neq \kb'. \E[\bmxb_{\kb}\bmxb_{\kb'}^H]=\bmzero$. Denote $\bmzb_{\kb}=\bmxb_{\kb}\otimes \bmhb_{\kb}$ which has zero mean and covariance
\beq
\begin{split}
    \bmC_{\bmzb_{\kb}}=\E[\bmzb_{\kb}\bmzb_{\kb}^H]=\sigma_x^2[\bmI_T\otimes(\bmC_{\bmhb_\kb|\bmuh}+\bmm_{\bmhb_\kb|\bmuh}\bmm_{\bmhb_{\kb}|\bmuh}^H)]
\end{split}
\eeq
Therefore, the likelihood can be represented by
\beq
    \begin{split}
         p(\bmY, \bmY_p|\bmUh)=p(\bmY|\bmY_p, \bmUh) p(\bmY_p|\bmUh)\\
         \simeq \mathcal{CN}(\bmy|\bmzero, \bmC_{\bmY|\bmY_p, \bmuh})\prod_{l, g}\mathcal{CN}(\bmyt_{p, lg}|\bmzero, \bmC_{\bmyt_{p, lg}|\bmuh})
    \end{split}
\eeq
where
\beq
\begin{split}
    \bmC_{\bmY|\bmY_p, \bmuh}=\bmC_v+\sum_{\kb\in U(\bmuh)}\bmC_{\bmzb_{\kb}}.
\end{split}
\eeq
The log-likelihood of $\bmuh$ is $\ln p(\bmY, \bmY_p|\bmUh)\simeq L_p(\bmuh)+L_d(\bmuh)$, where
\begin{align}
    &L_p(\bmuh)=\sum_{l, g}\ln \mathcal{CN}(\bmyt_{p, lg}|\bmzero, \bmC_{\bmyt_{p, lg}|\bmuh})\nonumber\\
    &=\sum_{l, g}-\ln[\det(\bmC_{\bmyt_{p, lg}|\bmuh})]-\bmyt_{p, lg}^H\bmC_{\bmyt_{p, lg}|\bmuh}^{-1}\bmyt_{p, lg}\label{eq:wsa2517}\\
        &L_d(\bmuh)=\ln \mathcal{CN}(\bmy|\bmzero, \bmC_{\bmY|\bmY_p, \bmuh})\nonumber\\
        &=-\ln[\det(\bmC_{\bmY|\bmY_p, \bmuh})]-\bmy^H\bmC_{\bmY|\bmY_p, \bmuh}^{-1}\bmy.\nonumber
\end{align}
Now we try to simplify $L_d$. Consider the large system limit, i.e., $L, \Kb, P, T\to \infty$ with fixed ratio between every pair. To have a realistic system, $[\bmY]_{n, t}$ and $\E\{|[\bmXb]_{\kb, t}|^2\}$ scale with $O(1)$. Therefore, $\tr[\bmXi_{\bmhb_{l\kb}}]$ scales with $O(1/K)$. This indicates that the off diagonal blocks of $\bmC_{\bmhb_\kb|\bmuh}+\bmm_{\bmhb_\kb|\bmuh}\bmm_{\bmhb_{\kb}|\bmuh}^H$ are higher-order perturbations of the blocks on the diagonal. We neglect the off diagonal blocks by approximating
\vspace{-1mm}
\beq
\vspace{-1mm}
    \forall l\neq l'. \sum_{\kb\in U(\bmuh)}\bmm_{\bmhb_{l\kb}|\bmuh}\bmm_{\bmhb_{l'\kb}|\bmuh}^H=\bmzero
\eeq
After that, $\bmC_{\bmhb_\kb|\bmuh}+\bmm_{\bmhb_\kb|\bmuh}\bmm_{\bmhb_{\kb}|\bmuh}^H$ is approximated as a block diagonal matrix with block size $N\times N$. This entails $\Lh_d\simeq L_d$ where
\beq
\vspace{-2mm}
\begin{split}
    \Lh_d(\bmuh)\!\!\simeq\!\! \sum_{l}\!\left\{\!-T\ln\!\left(\!\!\det\!\left[\!\bmC_v\!\!+\!\!\!\!\!\!\sum_{\kb\in U(\bmuh)}\!\!\!\!\!\sigma_x^2(\bmC_{\bmhb_{l\kb}|\bmuh}\!+\!\bmm_{\bmhb_{l\kb}}\bmm_{\bmhb_{l\kb}}^H)\right]\right)\right.\\
    \left.-\sum_{t}\bmy_{lt}^H[\bmC_v\!\!+\!\!\!\!\!\!\sum_{\kb\in U(\bmuh)}\!\!\!\!\sigma_x^2(\bmC_{\bmhb_{l\kb}|\bmuh}+\bmm_{\bmhb_{l\kb}|\bmuh}\bmm_{\bmhb_{l\kb}|\bmuh}^H)]^{-1}\bmy_{lt}\right\}
\end{split}\label{eq:wsa2516}
\eeq
where $\bmy_{lt}$ is the $t$-th column of $\bmY_l$. After the approximation, compared to $L_d(\bmuh)$, $\Lh_d(\bmuh)$ not only lowers the complexity, but also allows a distributed computation.

Combine both $L_p$ and approximated $\Lh_d$ and the likelihood $\bmuh$ can be finally approximated as
\beq
    L_{tot}(\bmuh)=\ln p(\bmY, \bmY_p|\bmUh)\simeq L_p(\bmuh)+\Lh_d(\bmuh).
    \label{eq:wsa2519}
\eeq

The constrained ML is NP-hard. Therefore, we present here an iterative method as a compromise. During the $i$-th iteration, we update the $u_{\kb}$ where $\kb=[(i-1) \text{ mod } \Kb]+1$ to maximize \eqref{eq:wsa2519}. We repeat this procedure until convergence.
The complete procedure is outlined in Algorithm~\ref{algo:wsa2501}, where \( \bme_{\kb} \) denotes a vector with a one in the \( \kb \)-th position and zeros elsewhere, and \( \bmE_{\kb} = \text{Diag}(\bme_{\kb}) \) is its corresponding diagonal matrix.

\floatstyle{spaceruled}
\restylefloat{algorithm}
\begin{algorithm}[t]
\caption{Iterative ML}\label{algo:wsa2501}
\begin{algorithmic}[1]
\State Initialize: $i=1$, $\bmuh=\bmzero$
\Repeat
\State $\kb= [(i-1) \text{ mod } \Kb]+1$
\State $\uh_k=\arg\max_{\theta\in \{0, 1\}}L_{tot}((\bmI-\bmE_{\kb})\bmuh+\bme_{\kb}\theta)$
\Until{Convergence}
\end{algorithmic}
\end{algorithm}

\section{Distributed VL-EP for Pseudo-Prior}
With estimated $\bmuh$ from previous section, we prune the estimated non-active users
\vspace{-1mm}
\beq
\vspace{-1mm}
    \bmH=\bmHb\bmUh;\; \bmH_l=\bmHb_l\bmUh;\; \bmX=\bmUh\bmXb;\; K=\bmuh^H\bmuh
\eeq
with $\bmh_k$, $\bmh_{lk}$, $\bmx_k^T$ as the $k$-th column of $\bmH$, $\bmH_l$ and the $k$-th row of $\bmX$ respectively. Furthermore, we use $\bmXi_{\bmh_{lk}}$, $G_g$ to denote a pruned version of $\bmXi_{\bmhb_{l\kb}}$ and $\Gb_g$.

Distributed VL-EP can be used for pseudo-prior of VB-EP. To estimate the $k$-th user channel, we consider other channels as interference and alternating estimate all the channels, i.e., minimize the negative log pdf $-\ln p(\bmY, \bmyt_{p, lg}, \bmh_{k})$ while treat $\forall k'\neq k. \bmh_{k'}$ as interference with distribution $\mathcal{CN}(\bmh_{k}'|\bmm_{\bmhc_{k'}}, \bmC_{\bmhc_{k'}})$.  
At each AP, 
the noise plus interference of received data $\bmy_l=\text{vec}(\bmY_l)$ has zero mean and covariance matrix
\beq
\bmC_{\bmvc_{lk}}\!\!=\!\sigma_v^2\bmI_{NT}\!+\!\!\!\sum_{k'\neq k}\!\!\sigma_{x}^2\bmI_T\otimes (\bmC_{\bmhc_{lk'}}\!\!\!\!+\!\bmm_{\bmhc_{lk}}\bmm_{\bmhc_{lk}}^H)
    \!=\!\bmI_T\otimes \bmC_{\bmvc_{lkt}}
\nonumber
\eeq
where
\beq
    \bmC_{\bmvc_{lkt}}=\sigma_v^2\bmI_N+\sum_{k'\neq k}\sigma_x^2(\bmC_{\bmhc_{lk'}}+\bmm_{\bmhc_{lk}}\bmm_{\bmhc_{lk}}^H) \label{eq:wsa2522}
\eeq
is constant in time since the noise power is constant in time and we define $\bmC_{\bmvc_{lk0}}=\bmC_{\bmvc_{lkt}}$.

On the other hand, if we look at the pilot part. The noise plus interference of the received pilot at each AP $\bmyt_{p, lG_g}$ relative to user $k$ can be modeled as Gaussian with the following mean and covariance matrix
\beq
\begin{split}
    \bmm_{\bmvc_{p, lk}}=\sum_{k'\in G_g/\{k\}}\bmm_{\bmhc_{lk'}};\; \bmC_{\bmvc_{p, lk}}=\bmC_{\bmvt}+\sum_{k'\in G_g/\{k\}}\bmC_{\bmhc_{lk'}}\label{eq:wsa2523}
\end{split}
\eeq

In order to construct a majorization function, we first look at the approximated marginal joint pdf (with interference approximated as Gaussian noise)
\begin{align}
    &p(\bmY, \bmyt_{p, lg}, \bmh_k, \bmx_k)=p(\bmY|\bmh_k, \bmx_k) p(\bmyt_{p, lg}|\bmh_k)p(\bmh_k)p(\bmx_k)\nonumber\\
    &=p(\bmx_k)\prod_{l}p(\bmY_l|\bmh_{lk}, \bmx_{k})p(\bmyt_{p, lg}|\bmh_{lk})p(\bmh_{lk})\label{eq:wsa2530p}
\end{align}
The effect term of the majorization function at $\bmh_k=\bmhc_{k}$ is \cite{9723203}
\beq
    \text{em}(\bmh_k|\bmHc)=-\E_{\bmx_k|\bmY, \bmm_{\bmhc_{k}}}[\ln p(\bmY, \bmyt_{p, lg}, \bmh_k, \bmx_k)].
    \label{eq:wsa2534}
\eeq
The channel $\bmh_k$ is estimated by $\arg\min_{\bmh_k} \text{em}(\bmh_k|\bmHc)$ iteratively.
\subsection{Expectation Step}
To minimize the majorization function \eqref{eq:wsa2534}, we first need to get the posterior $p(\bmx_k|\bmY, \bmm_{\bmhc_{k}})$. Rewrite it into the product of prior and likelihood
\beq
    p(\bmx_k|\bmY, \bmm_{\bmhc_{k}})\propto p(\bmx_k) \prod_l p(\bmY_l|\bmx_k, \bmm_{\bmhc_{lk}}) \label{eq:Laasdfsda}
\eeq
We observe that the likelihood of the observation at the $l$-th AP, which is proportional to Gaussian distribution:
\beq
    p(\bmY_l|\bmx_k, \bmm_{\bmhc_{lk}})\propto \mathcal{CN}(\bmx_k|\bmm_{\bmxd_{lk}}, \bmC_{\bmxd_{lk}}),
\eeq
where
\beq
    \begin{split}
        &\bmC_{\bmxd_{lk}}=\tau_{\bmxd_{lk}}^{i}\bmI_T ;\; m_{\xd_{lkt}}=\tau_{\bmxd_{lk}}^{i} \bmm_{\bmhc_{lk}}^{H}\bmC_{\bmvc_{lk0}}^{-1}\bmy_{lt};\\
        &\tau_{\bmxd_{lk}}=\left(\bmm_{\bmhh_{lk}}^{(i-1) H}\bmC_{\bmvb_{lk}}^{-1}\bmm_{\bmhh_{lk}}^{(i-1)}\right)^{-1}
    \end{split}
    \label{eq:wsa2528}
\eeq
Since $\bmC_{\bmxd_{lk}}$ is a diagonal matrix, we can compute the posterior statistics of $x_{kt}$ by one-dimensional integrals. We denote marginalized version of \eqref{eq:Laasdfsda}
\beq
\begin{split}
    \pc(x_{kt})=\int p(\bmx_k|\bmY_l, \bmm_{\bmhc_{lk}}) dx_{\{k'\neq k\}\{t'\neq t\}}\\
    =p(x_{kt})\prod_l \mathcal{CN}(x_{kt}| m_{\xd_{lkt}}, \tau_{\bmxd_{lk}})
\end{split}
\eeq
without the pseudo-prior, using discrete $p(x_{kt})$ may lead to convergence problem. Therefore, we approximate $p(x_{kt})$ by Gaussian $\mathcal{CN}(x_{kt}|0, \sigma_x^2)$. As a result, the approximated mean and variance are
\beq
\begin{split}
    \tau_{\xc_{kt}}=(\sigma_x^{-2}+\sum_{l}\tau_{\bmxd_{lk}}^{-1})^{-1};\; m_{\xc_{kt}}=\tau_{\xc_{kt}}\sum_{l}\tau_{\bmxd_{lk}}^{-1}m_{\xd_{lkt}}
\end{split}
\label{eq:wsa2530}
\eeq
To estimate the posterior $m_{\xc_{kt}}$, 
We denote the posterior mean, variance and second-order moment as $m_{\xc_{kt}}$, $\tau_{\xc_{kt}}$ and $r_{\xc_{kt}}$.

\subsection{Maximization Step}
Investigate the log pdf inside the expectation in \eqref{eq:wsa2534}. By neglecting terms that do not contain $\bmh_{lk}$
\beq
\begin{split}
    &-\ln p(\bmY, \bmyt_{p, lg}, \bmh_k, \bmx_k)\\
    &=\sum_l \left[(\bmx_k\otimes \bmI_N)\bmh_{lk}-\bmy_l\right]^H\bmC_{\bmvh_{lk}}^{-1}\left[(\bmx_k\otimes \bmI_N)\bmh_{lk}-\bmy_l\right]\\
    &+(\bmh_{lk}-\bmyt_{p, lg}+\bmm_{\bmvh_{p, lk}})^H\bmC_{\bmvh_{p, lk}}^{-1}(\bmh_{lk}-\bmyt_{p, lg}+\bmm_{\bmvh_{p, lk}})\\
    &+\bmh_{lk}^H\bmXi_{\bmh_{lk}}^{-1}\bmh_{lk}+\text{const.}
\end{split}
\eeq
Therefore, $\arg\min_{\bmh_{lk}} \text{em}(\bmh_{lk}|\bmHc)$ leads to the updates
\beq
\begin{split}
    &\bmC_{\bmhc_{lk}}=\left(\bmC_{\bmvc_{p, lk}}^{-1}+\bmC_{\bmvc_{lk0}}^{-1}\sum_t r_{\xc_{kt}}+\bmXi_{\bmh_{lk}}^{-1}\right)^{-1}\\
    &\bmm_{\bmhc_{lk}}=\bmC_{\bmhc_{lk}}\left[\sum_t m_{\xc_{kt}}^*\bmC_{\bmvc_{lk0}}^{-1}\bmy_{lt}+\bmC_{\bmvc_{p, lk}}^{-1}(\bmyt_{p, lg}-\bmm_{\bmvc_{p, lk}})\right]
\end{split}
\label{eq:wsa2532}
\eeq
The above procedure is illustrated in Algorithm \ref{algo:wsa2502}. To make the algorithm robust, we change the order of maximization step and expectation step. 

\floatstyle{spaceruled}
\restylefloat{algorithm}
\begin{algorithm}[t]
\caption{VL-EP For Pseudo Prior}\label{algo:wsa2502}
\begin{algorithmic}[1]
\State Initialize: $\tau_{\xc_{kt}}=1$, $m_{\xc_{kt}}=0$
\Repeat
\State $\bmC_{\bmhc_{lk}}$, $\bmm_{\bmhc_{lk}}$ by \eqref{eq:wsa2532}

\State $\bmC_{\bmvc_{lk0}}$, $\bmm_{\bmvc_{p, lk}}$, $\bmC_{\bmvc_{p, lk}}$ by \eqref{eq:wsa2522}, \eqref{eq:wsa2523}

\State $\tau_{\xc_{kt}}$, $m_{\xc_{kt}}$ by \eqref{eq:wsa2528}, \eqref{eq:wsa2530}

\Until{Converge}
\end{algorithmic}
\end{algorithm}

\section{Bilinear VB-EP}
Inspired by \cite{pseudoprior}, we then integrate the pseudo prior from VL-EP to the prior distribution of $\bmh_{lk}$ and $\bmx_k$ to form a enhanced prior
\beq
\begin{split}
    &\pbv(\bmh_{lk})=\mathcal{CN}(\bmh_{lk}|\bmm_{\bmhbv_{lk}}, \bmC_{\bmhbv_{lk}})\\
    &:=\mathcal{CN}(\bmh_{lk}|\bmm_{\bmhc_{lk}}, \bmC_{\bmhc_{lk}})\mathcal{CN}(\bmh_{lk}|\bmzero, \bmXi_{\bmh_{lk}})\\
    &\pbv(x_{kt})=\pc(x_{kt})p(x_{kt})
\end{split}
\eeq
The additional potentials do not contribute additional information, they only serve as regularization to prevent VB-EP from drifting away from optimal solution. Furthermore, we can define the pilot-based enhanced prior as
$
    \ptbv(\bmh_{lG_g})\propto p(\bmyt_{p, lg}|\bmh_{lG_g})\pbv(\bmh_{lG_g}).
$
If we introduce $\forall k. \bmZ_{lk}=\bmh_{lk}\bmx_{k}^T$ and $\bmz_{lk}=\text{vec}(\bmZ_{lk})$, we can verify that our current system model has exactly the same form of \cite{VBEP}. The VB-EP algorithm aims to find local approximation of the marginal posterior (also called belief, denoted by $b_{\cdot}(\cdot)$) with  messages $\bmmu_{\cdot;\cdot}(\cdot)$ of their neighboring variables. From \cite{VBEP}, these approximations can be written as
\vspace{-1mm}
\begin{align}
\vspace{-1mm}
    &b_{f_{\bmz_l}}(\bmz_{l\{k\}})\!=\!p(\bmY_l|\bmz_{l\{k\}})\!\prod_{k}\!\mu_{\bmz_{lk}; f_{\bmz_l}}(\bmz_{lk})\nonumber\\
    &b_{f_{\bmh_{lG_g}}}(\bmh_{lG_g})=\ptbv( \bmh_{lG_g})\prod_{k\in G_g}\mu_{\bmh_{lk};f_{\bmh_{lG_g}}}(\bmh_{lk})\nonumber\\
    &b_{f_{\bmx_k}}(\bmx_k)=\pbv(\bmx_{k})\mu_{\bmx_k;f_{\bmx_k}}(\bmx_k)\label{eq:wsa2536}\\
    &b_{\delta_{\bmh, lk}}\!(\bmh_{lk})\!\!=\!\mu_{\bmh_{lk};\delta_{lk}}(\bmh_{lk})e^{\int \!\!b_{\delta_{\bmx, lk}}\!(\bmx_k)\!\ln \mu_{\bmz_{lk};\delta_{lk}}(\text{vec}(\bmh_{lk}\bmx_k^{\mathsf{T}}))d\bmx_k}\nonumber\\
    &b_{\delta_{\bmx, lk}}\!(\bmx_k)\!=\!\mu_{\bmx_k;\delta_{lk}}\!(\bmx_k)e^{\int \! b_{\delta_{\bmh, lk}}\!(\bmh_{lk})\ln \mu_{\bmz_{lk};\delta_{lk}}(\text{vec}(\bmh_{lk}\bmx_k^{\mathsf{T}}))d\bmh_{lk}}\nonumber\\
    &b_{\bmz_{lk}}(\bmz_{lk})=\mu_{\bmz_{lk};f_{\bmz_l}}(\bmz_{lk})\mu_{\bmz_{lk};\delta_{lk}}(\bmz_{lk})\nonumber\\
    &b_{\bmh_{lk}}(\bmh_{lk})=\mu_{\bmh_{lk};f_{\bmh_{lG_g}}}(\bmh_{lk})\mu_{\bmh_{lk};\delta_{lk}}(\bmh_{lk})\nonumber\\
    &b_{\bmx_{k}}(\bmx_k)=[\mu_{\bmx_k;f_{\bmx_k}}(\bmx_k)\prod_l\mu_{\bmx_k;\delta_{lk}}(\bmx_k)]^{1/L}\nonumber,
\end{align}
where we use $f_{\bmz_l}$, $f_{\bmx_k}$ $f_{\bmh_{lG_g}}$ to denote the likelihood of $p(\bmy_l|\bmz_{l\{k\}})$, the enhanced prior $\pbv(\bmx_k)$ and the pilot based enhanced prior $\ptbv(\bmh_{lG_g})$.
Due to the definition of $\bmZ_{lk}$, the likelihood of $\bmZ_{lk}$ given $\bmh_{lk}\bmx_{k}^T$ is a Dirac delta function. Therefore, in the equations above, we use $\delta_{lk}$ to denote $p(\bmZ_{lk}|\bmh_{lk}\bmx_{k}^T)$.
The messages are the solution system of equations consisting of the following moments constraints
\begin{align}
        &\E_{b_{f_{\bmz_l}}}[\phi_{\bmz_{lk}}(\bmz_{lk})]=\E_{b_{\bmz_{lk}}}[\phi_{\bmz_{lk}}(\bmz_{lk})]\label{eq:cons1}\\
        &\E_{\delta_{lk}}[\phi_{\bmz_{lk}}(\bmz_{lk})]=\E_{b_{\bmz_{lk}}}[\phi_{\bmz_{lk}}(\bmz_{lk})]\label{eq:cons2}\\
        &\E_{b_{f_{\bmh_{lG_g}}}}[\phi_{\bmh_{lk}}(\bmh_{lk})]=\E_{b_{\bmh_{lk}}}[\phi_{\bmh_{lk}}(\bmh_{lk})]\label{eq:cons3}\\
        &\E_{b_{\delta_{lk}}}[\phi_{\bmh_{lk}}(\bmh_{lk})]=\E_{b_{\bmh_{lk}}}[\phi_{\bmh_{lk}}(\bmh_{lk})]\label{eq:cons4}
\end{align}
and the following marginal constraints
\begin{align}
        &\int b_{\delta_{lk}}(\bmz_{lk}, \bmh_{lk}, \bmx_k) d\bmz_{lk} d\bmh_{lk}=b_{\bmx_k}(\bmx_k)\label{eq:consx1}\\
        &b_{f_{\bmx_k}}(\bmx_k)=b_{\bmx_k}(\bmx_k).\label{eq:consx2}\\
        &b_{\delta_{lk}}(\bmz_{lk}, \bmh_{lk}, \bmx_k)\!=\!b_{\delta_{\bmh,lk}}(\bmh_{lk})b_{\delta_{\bmx, lk}}(\bmx_k)\delta(\bmZ_{lk}\!-\!\bmh_{lk}\bmx_k^{\mathsf{T}}).\label{eq:icassp248}
\end{align}
where $\phi_{\cdot}(\cdot)$ denotes the first and second order moments. For all the factors $f$ and variables $\theta$, we say they are neighboring to each other if $\theta$ is an argument of $f$. The messages describing the beliefs in \eqref{eq:wsa2536} are all of the form $\mu_{\theta;f}$ which can be interpreted as messages from variable to factor. However, to make the derivation easier, we define factor to variable messages $\mu_{f;\theta}$ such that
\beq
    \forall f\in N(\theta), \mu_{\theta;f}(\theta)=\!\!\!\!\prod_{f'\in N(\theta)/\{f\}}\!\!\!\!\mu_{f';\theta}(\theta),\label{eq:icassp23}
\eeq
where $N(\theta)$ denotes the neighborhood around $\theta$. As we can see, group of factor-to-variable messages is isomorphic to the group of varible-to-factor messages. Therefore, in the following context, we only describe the iterative update of factor-to-variable messages which can also be understood as the feedback messages from the factors.

\section{Iterative Solution to VB-EP}
By considering \eqref{eq:consx2}, we get $\mu_{f_{\bmx_k}}=\pbv(\bmx_k)$. Following a similar approach, we can get the following non-trivial updates.
\subsection{Update of $\mu_{f_{\bmz_l};\bmz_{lk}}$}
By making \eqref{eq:cons1} consistent while keeping all the message except for $\mu_{f_{\bmz_l};\bmz_{lk}}=\mu_{\bmz_{lk};\delta_{\bmz_l}}$ fixed, we can obtain the update for 
$\bmmu_{f_{\bmz_{l}};\bmz_{lk}}$.
\vspace{-2mm}
\beq
\vspace{-2mm}
    \mu_{f_{\bmz_l};\bmz_{lk}}(\bmz_{lk})=\mathcal{CN}(\bmz_{lk}|\bmy_l-\!\!\sum_{k'\neq k}\!\!\bmm_{\bmz_{lk'};f_{\bmz_l}}, \bmC_\bmv\!+\!\!\sum_{k'\neq k}\bmC_{\bmz_{lk'};f_{\bmz_l}}).\label{eq:wsa2554}
\eeq

\subsection{Update of $\mu_{f_{\bmH_{lG_g}};\bmh_{lk}}$}
The consistency of \eqref{eq:cons3} result to the update of $\mu_{f_{\bmH_{lG_g}};\bmh_{lk}}=\mu_{\bmh_{lk};\delta_{lk}}$ by considering all the other messages fixed.
\beq
    \begin{split}
        &b_{f_{\bmH_{lG_g}}}(\bmH_{lG_g})=p(\bmyt_{p, lg}|\sum_{k\in G_g}\bmh_{lk}, \bmC_{\vt})\\
        &\cdot\prod_{k\in G_g}\pbv(\bmh_{lk})\mu_{\bmh_{lk};f_{\bmH_{lG_g}}}(\bmh_{lk}); \text{ and}\\
        &\mu_{f_{\bmH_{lG_g}};\bmh_{lk}}(\bmh_{lk})=\frac{\int b_{f_{\bmH_{lG_g}}}(\bmh_{lG_g}) d\bmh_{l\{k'\neq k\}}}{\mu_{\bmh_{lk};f_{\bmH_{lG_g}}}(\bmh_{lk})}
    \end{split}
\eeq
Based on Gaussian reproduction lemma, we have
\beq
\begin{split}
    \mu_{f_{\bmH_{lG_g}};\bmh_{lk}}(\bmh_{lk})=\mathcal{CN}\left(\bmh_{lk}| \bmm_{\bmY_p|\bmh_{lk}}, \bmC_{\bmY_p|\bmh_{lk}}\right)\pc(\bmh_{lk})
\end{split}
\eeq
where the approximated likelihood is captured by
\vspace{-2mm}
\beq
    \begin{split}
        &\bmC_{\bmY_p|\bmh_{lk}}=\bmC_{\vt}+\sum_{k'\in G_g/\{k\}}\bmC_{\bmht_{lk'}}\\
        &\bmm_{\bmY_p|\bmh_{lk}}=\bmyt_{p, lg}-\sum_{k'\in G_g/\{k\}}\bmm_{\bmht_{lk'}}\\
    &\bmC_{\bmht_{lk}}=\left(\bmC_{\bmh_{lk};f_{\bmH_{lG_g}}}^{-1}+\bmC_{\bmhbv_{lk}}^{-1}\right)^{-1}\\
    &\bmm_{\bmht_{lk}}=\bmC_{\bmht_{lk}}\left(\bmC_{\bmh_{lk};f_{\bmH_{lG_g}}}^{-1}\bmm_{\bmh_{lk};f_{\bmH_{lG_g}}}+\bmC_{\bmhbv_{lk}}^{-1}\bmm_{\bmhbv_{lk}}\right)
\end{split}
\label{eq:wsa2557}
\eeq
The message mean and covariance is obtained as

\begin{align}
    &\bmC_{f_{\bmH_{lG_g}};\bmh_{lk}}=\left(\bmC_{\bmY_p|\bmh_{lk}}^{-1}+\bmC_{\bmhbv_{lk}}^{-1}\right)^{-1}\label{eq:wsa2558}\\
    &\bmm_{f_{\bmH_{lG_g}};\bmh_{lk}}\!\!=\!\bmC_{f_{\bmH_{lG_g}};\bmh_{lk}}\left(\bmC_{\bmY_p|\bmh_{lk}}^{-1}\bmm_{\bmY_p|\bmh_{lk}}\!\!\!+\!\bmC_{\bmhbv_{lk}}^{-1}\bmm_{\bmhbv_{lk}}\right)\nonumber
\end{align}

\subsection{Update of $\mu_{\delta_{lk};\bmx_{k}}$}
According to \eqref{eq:consx1}, the feedback message is
\beq
\vspace{-2mm}
    \mu_{\delta_{lk};\bmx_{k}}(\bmx_{k})\propto \prod_t \mathcal{CN}(x_{kt}|\widehat{m}_{\delta_{lk};x_{kt}}, \widehat{\tau}_{\delta_{lk};x_{kt}}), \label{eq:icassp2529}
\eeq
with
\vspace{-1mm}
\beq
\begin{split}
    &\tau_{\delta_{lk};x_{kt}}=\tr\left[\{\bmC_{\bmz_{lk};\delta_{lk}}\}_{tt, N}^{-1}\bmR_{b_{\delta_{\bmh, lk}}}\right]^{-1}\\
    &m_{\delta_{lk};x_{kt}}\!\!=\widehat{\tau}_{\delta_{lk};x_{kt}}\bmm_{b_{\delta_{\bmh, lk}}}^{\mathsf{H}}\{\bmC_{\bmz_{lk};\delta_{lk}}\}_{tt, N}^{-1}\{\bmm_{\bmz_{lk};\delta_{lk}}\}_{t, N},
\end{split}
\label{eq:wsa2560}
\eeq
where $\bmm_{b_{\delta_{\bmh, lk}}}$ and $\bmR_{b_{\delta_{\bmh, lk}}}=\bmC_{b_{\delta_{\bmh, lk}}}+\bmm_{b_{\delta_{\bmh, lk}}}\bmm_{b_{\delta_{\bmh, lk}}}^\mathsf{H}$ denote the mean and correlation matrix of the Gaussian pdf $b_{\delta_{\bmh, lk}}$ and the notation $\{\cdot\}_{t_1t_2, N}$ denotes the block entry of size $N\times N$ located in the $t_1$-th block row and $t_2$-th block column. From this point, we can also update the belief by 
\beq
b_{\delta_{\bmx, lk}}(\bmx_k)=\mu_{\delta_{lk};\bmx_{k}}(\bmx_k)\mu_{\bmx_{k};\delta_{lk}}(\bmx_k).\label{eq:icassp2532}
\eeq

\subsection{Update of $\mu_{\delta_{lk};\bmh_{lk}}$}
Satisfying the consistency constraint \eqref{eq:cons4} by treating $\mu_{\delta_{lk};\bmh_{lk}}=\mu_{\bmh_{lk};f_{\bmH_{lG_g}}}$ as the only variable and we can obtain the update equation for it. The belief $b_{\delta_{\bmh, lk}}$ in \eqref{eq:wsa2536} can be verified as Gaussian since the exponential term is quadratic. Therefore, the outbound message is computed by
$\mu_{\delta_{lk};\bmh_{lk}}(\bmh_{lk})=b_{\delta_{\bmh, lk}}(\bmh_{lk})/\mu_{\bmh_{lk};\delta_{lk}}(\bmh_{lk})$. The mean and covariance matrix of the Gaussian message $\mu_{\delta_{lk};\bmh_{lk}}$ are 
\begin{align}
    &\bmC_{\delta_{lk};\bmh_{lk}}=\left(\sum_t [\bmr_{b_{\delta_{\bmx, lk}}}]_t\{\bmC_{\bmz_{lk};\delta_{lk}}\}_{tt, N}^{-1}\right)^{-1}\label{eq:wsa2562}\\
    &\bmm_{\delta_{lk};\bmh_{lk}}\!\!\!\!=\!\! \bmC_{\delta_{lk};\bmh_{lk}}\!\!\!\left(\!\!\sum_{t} [\bmm_{b_{\delta_{\bmx, lk}}}]_{t}^*\{\bmC_{\bmz_{lk};\delta_{lk}}\}_{tt, N}\{\bmm_{\bmz_{lk};\delta_{lk}}\}_{t, N}\!\!\!\right),\nonumber
\end{align}
where
$
\bmr_{b_{\delta_{\bmx, lk}}}=\E_{b_{\delta_{\bmx, lk}}}[\bmx_k\bmdot\bmx_k^*], \,\bmm_{b_{\delta_{\bmx, lk}}}=\E_{b_{\delta_{\bmx, lk}}}[\bmx_k]
$
with "$\bmdot$" denoting element-wise product. We update the belief by 
\beq
\begin{split}
b_{\delta_{\bmh, lk}}(\bmh_{lk})=\mathcal{CN}({\bmh_{lk}|\bmm_{b_{\delta_{\bmh, lk}}}, \bmC_{b_{\delta_{\bmh, lk}}}})\\
=\mu_{\delta_{lk};\bmh_{lk}}(\bmh_{lk})\mu_{\bmh_{lk};\delta_{lk}}(\bmh_{lk}).\label{eq:icassp2533}
\end{split}
\eeq

\subsection{Update of $\mu_{\delta_{lk};\bmz_{lk}}$ \label{SectionF}}

We examine the moments constraint given by \eqref{eq:cons2}. Based on the additional constraint \eqref{eq:icassp248}, the update equation of $\mu_{\delta_{lk};\bmz_{lk}}$ can be derived as
\vspace{-1mm}
\beq
\vspace{-1mm}
    \mu_{\delta_{lk};\bmz_{lk}}(\bmz_{lk})=\frac{\text{proj}[b_{\delta_{lk}}(\bmz_{lk})]}{\mu_{\bmz_{lk};\delta_{lk}}(\bmz_{lk})},
\eeq
where the operation $q(\bmz_{lk})=\text{proj}[p(\bmz_{lk})]$ projects the distribution $p$ to Gaussian family $q$ of block covariance matrices, such that the sufficient statistics $\phi_{\bmz_{lk}}(\bmz_{lk})$ of $p$ and $q$ are the same. Thus, the message $\mu_{\delta_{lk};\bmz_{lk}}$ is updated by 
\beq
\begin{split}
    \mu_{\delta_{lk};\bmz_{lk}}(\bmz_{lk})=\mathcal{CN}(\bmz_{lk}|\bmm_{\delta_{lk};\bmz_{lk}},\bmC_{\delta_{lk};\bmz_{lk}})\\
    =\frac{\mathcal{CN}(\bmz_{lk}|\bmm_{b_{\delta_{\bmz,lk}}}, \bmC_{b_{\delta_{\bmz,lk}}})}{\mathcal{CN}(\bmz_{lk}|\bmm_{\bmz_{lk};\delta_{lk}}, \bmC_{\bmz_{lk}; \delta_{lk}})},
\end{split}
\label{eq:wsa2565}
\eeq
with
\begin{align}
    &\bmm_{b_{\delta_{\bmz,lk}}}\!\!\!=\bmm_{b_{\delta_{\bmx, lk}}}\otimes \bmm_{b_{\delta_{\bmh, lk}}}\nonumber\\
    &\bmC_{b_{\delta_{\bmz,lk}}}\!\!\!=\text{diag}(\bmr_{b_{\delta_{\bmx, lk}}})\otimes \bmC_{b_{\delta_{\bmh, lk}}}\!\!\!+\bmC_{b_{\delta_{\bmx, lk}}}\otimes \bmm_{b_{\delta_{\bmh, lk}}}\bmm_{b_{\delta_{\bmh, lk}}}^\mathsf{H},\nonumber
\end{align}
where $\bmC_{b_{\delta_{\bmx, lk}}}=\E_{b_{\delta_{\bmx, lk}}}[(\bmx_k-\bmm_{b_{\delta_{\bmx, lk}}})(\bmx_k-\bmm_{b_{\delta_{\bmx, lk}}})^{\mathsf{H}}]$ is the covariance of the belief $b_{\delta_{\bmx, lk}}$. The above procedure is concluded in \textbf{Algorithm} \ref{algo:wsa2503}.
Finally, by combining all the previous section together, we sum up the whole algorithm in \textbf{Algorithm} \ref{algo:iccasp241}.

\floatstyle{spaceruled}
\restylefloat{algorithm}
\begin{algorithm}[t]
\caption{VB-EP \cite{VBEP}}\label{algo:wsa2503}
\begin{algorithmic}[1]
\State Initialize: all factor-to-variable messages 
\Repeat
\State $\bmC_{f_{\bmz_l};\bmz_{lk}}$, $\bmm_{f_{\bmz_l};\bmz_{lk}}$ by \eqref{eq:wsa2554}
\State $\bmC_{f_{\bmH_{lG_g}};\bmh_{lk}}$, $\bmm_{f_{\bmH_{lG_g}};\bmh_{lk}}$ by \eqref{eq:wsa2557}-\eqref{eq:wsa2558}

\State $b_{\delta_{\bmh, lk}}(\bmh_{lk})=\mu_{\bmh_{lk};\delta_{lk}}(\bmh_{lk})\mu_{\delta_{lk};\bmh_{lk}}(\bmh_{lk})$
\State $\tau_{\delta_{lk};x_{kt}}$, $m_{\delta_{lk};x_{kt}}$ by \eqref{eq:wsa2560}

\State $\mu_{\bmx_{k};\delta_{lk}}(\bmx_k)=\pc(\bmx_k)\prod_{l'\neq l}\mu_{\delta_{l'k};\bmx_{k}}(\bmx_k)$

\State $b_{\delta_{\bmx, lk}}(\bmx_k)=\mu_{\bmx_{k};\delta_{lk}}(\bmx_k)\mu_{\delta_{lk};\bmx_{k}}(\bmx_k)$

\State $\bmC_{\delta_{lk};\bmh_{lk}}$, $\bmm_{\delta_{lk};\bmh_{lk}}$ by \eqref{eq:wsa2562}

\State $b_{\delta_{\bmh, lk}}(\bmh_{lk})=\mu_{\bmh_{lk};\delta_{lk}}(\bmh_{lk})\mu_{\delta_{lk};\bmh_{lk}}(\bmh_{lk})$

\State $\bmC_{\delta_{lk};\bmz_{lk}}$, $\bmm_{\delta_{lk};\bmz_{lk}}$ by \eqref{eq:wsa2565}

\Until{Converge}
\end{algorithmic}
\end{algorithm}

\floatstyle{spaceruled}
\restylefloat{algorithm}
\begin{algorithm}[t]
\caption{Proposed Method}\label{algo:iccasp241}
\begin{algorithmic}[1]
\State [Initialize $\bmU$]$\Rightarrow$Algorithm \ref{algo:wsa2501}

\State $(K, G_g, \bmXi_{\bmh_{lk}})=\text{prune}(\Kb, \Gb_g, \bmXi_{\bmhb_{l\kb}})$
\State [Pseudo Prior]$\Rightarrow$Algorithm \ref{algo:wsa2502}

\State [VB-EP]$\Rightarrow$Algorithm \ref{algo:wsa2503}

\end{algorithmic}
\end{algorithm}

\section{Simulation Results}
We consider a CF MaMIMO system covering a $400\times 400 m^2$ area. A total of $16$ APs are located at coordinates $\{\frac{400i}{3}+\frac{400j}{3}|i,j\in[0, 3]\}$ with a height of $10m$. The $\Kb=16$ UTs are uniformly distributed in the area. To evaluate the performance of the algorithm, we generate $50$ UT location setups. Following the settings in \cite{10694527, karataev2024bilinear}, the data symbol power is set to $\sigma_x^2=14 \text{dbm}$, the noise power to $\sigma_v^2=-96 \text{dbm}$ the $u_k$ are generated by uniform distribution. The channel fading model is defined based on the distance $d_{lk}$ between AP $l$ and UT $\kb$ as
$
    \sigma_{h_{lk}}^2=-30.5-36.7\log_{10}\left(\frac{d_{lk}}{1\text{m}}\right) \text{dB}
$. We set $P=8$ with orthogonal pilot sequence randomly assigned to two users. The data sequence contains $4$QAM symbols and has length $T=20$.
For each UT position setup, we generate $50$ iterations. Consider three performance metrics, Channel Normalized Mean Squared Error(CNMSE), Activity Detection Error Rate (DER) and Symbol Error Rate (SER), For each metric, we compute the empirical cumulative distribution function (CDF) across the different position setups. We compare the proposed PP-VB-EP with the "Joint Activity
Detection, Channel Estimation, and Data Detection based
on expectation propagation" (JACD-EP) presented in \cite{10694527}. The results of our proposed method is ploted with solid curved while the ones from JACD-EP are plotted with dotted curves. The Algorithm 1 is able to give a better result compared to previous work, which verifies the correctness of our approximated log-likelihood \eqref{eq:wsa2519}. However, since we neglect the estimation error of $\bmU$, the estimation of $\bmH$ and $\bmX$ performs slightly worse than the previous algorithm.

\begin{figure}[t]
    \centering
    \includegraphics[width=0.485\textwidth]{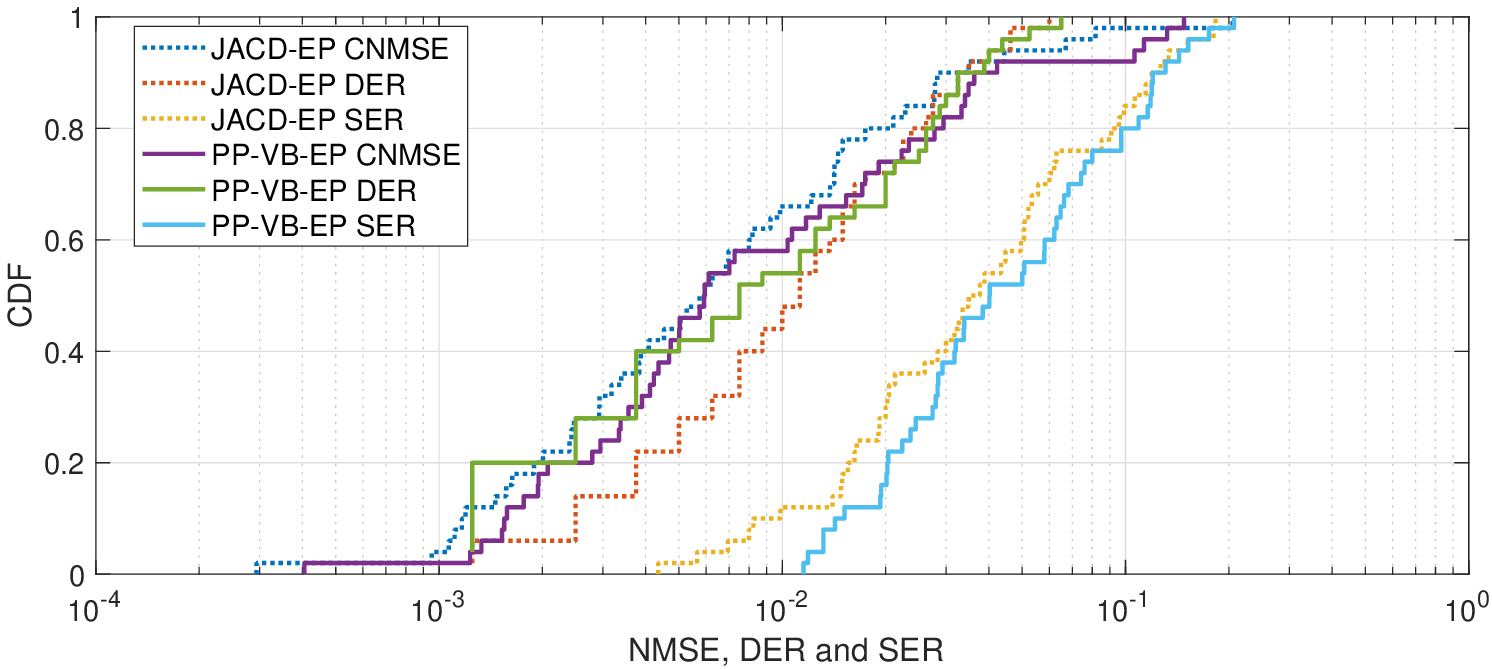}
    \vspace{-6mm}
    \caption{Comparison between PP-VB-EP and JACD-EP} 
    \vspace{-4mm}
    \label{fig:N}
    \vspace{-2mm}
\end{figure}

\section{Conclusions}

We propose an iterative distributed Maximum Likelihood (ML) algorithm for activity detection, which simplifies the joint probability density function \( p(\bmY, \bmY_p) \) by leveraging the CLT and large-system approximations. Building on the concept of pseudo priors, we introduce PP-VB-EP, which incorporates a distributed version of the VL-EP algorithm. Simulation results demonstrate that our proposed classical iterative distributed ML algorithm outperforms the EP-based approach in activity detection. However, due to the neglecting of detection errors of user activity, some performance degradation is observed in channel estimation and data detection. Nevertheless, VB-EP allows us to handle continuous data, and notably, its computational complexity does not scale with the constellation size.

{\bf Acknowledgements}
EURECOM's research is partially supported by its industrial members:
ORANGE, BMW, SAP, iABG,  Norton LifeLock, by the Franco-German project 
CellFree6G (ANR),
the French project PERSEUS (PEPR-5G), the EU H2030 project CONVERGE, and
by a Huawei France funded Chair towards Future Wireless Networks.

\bibliographystyle{IEEEtran}

\bibliography{wsa25_ref}

\begin{thebibliography}{1}
\providecommand{\url}[1]{#1}
\csname url@samestyle\endcsname
\providecommand{\newblock}{\relax}
\providecommand{\bibinfo}[2]{#2}
\providecommand{\BIBentrySTDinterwordspacing}{\spaceskip=0pt\relax}
\providecommand{\BIBentryALTinterwordstretchfactor}{4}
\providecommand{\BIBentryALTinterwordspacing}{\spaceskip=\fontdimen2\font plus
\BIBentryALTinterwordstretchfactor\fontdimen3\font minus \fontdimen4\font\relax}
\providecommand{\BIBforeignlanguage}[2]{{%
\expandafter\ifx\csname l@#1\endcsname\relax
\typeout{** WARNING: IEEEtran.bst: No hyphenation pattern has been}%
\typeout{** loaded for the language `#1'. Using the pattern for}%
\typeout{** the default language instead.}%
\else
\language=\csname l@#1\endcsname
\fi
#2}}
\providecommand{\BIBdecl}{\relax}
\BIBdecl

\bibitem{7827017}
H.~Q. Ngo, A.~Ashikhmin, H.~Yang, E.~G. Larsson, and T.~L. Marzetta, ``{Cell-Free Massive MIMO Versus Small Cells},'' \emph{IEEE Transactions on Wireless Communications}, 2017.

\bibitem{9723203}
R.~Gholami, L.~Cottatellucci, and D.~Slock, ``{Message Passing for a Bayesian Semi-Blind Approach to Cell-Free Massive MIMO},'' in \emph{Asilomar Conference on Signals, Systems, and Computers}, 2021.

\bibitem{10694527}
C.~Forsch, A.~Karataev, and L.~Cottatellucci, ``{Distributed Joint User Activity Detection, Channel Estimation, and Data Detection via Expectation Propagation in Cell-Free Massive MIMO},'' in \emph{IEEE International Workshop on Signal Processing Advances in Wireless Communications (SPAWC)}, 2024.

\bibitem{karataev2024bilinear}
A.~Karataev, C.~Forsch, and L.~Cottatellucci, ``{Bilinear Expectation Propagation for Distributed Semi-Blind Joint Channel Estimation and Data Detection in Cell-Free Massive MIMO},'' \emph{IEEE Open Journal of Signal Processing}, 2024.

\bibitem{8454392}
L.~Liu, E.~G. Larsson, W.~Yu, P.~Popovski, C.~Stefanovic, and E.~de~Carvalho, ``{Sparse Signal Processing for Grant-Free Massive Connectivity: A Future Paradigm for Random Access Protocols in the Internet of Things},'' \emph{IEEE Signal Processing Magazine}, 2018.

\bibitem{EURECOM+7308}
Z.~Zhao and D.~Slock, ``{Semi-Blind Sparse Channel Learning in Cell-Free Massive MIMO - a CRB Analysis},'' in \emph{ICC, International Conference on Communications}, IEEE, Ed., 2023.

\bibitem{VBEP}
------, ``{Decentralized Message-passing for Semi-Blind Channel Estimation in Cell-Free Systems Based on Bethe Free Energy Optimization},'' in \emph{Asilomar Conference on Signals, Systems, and Computers}, 2024.

\bibitem{9351786}
D.~Zhang, X.~Song, W.~Wang, G.~Fettweis, and X.~Gao, ``{Unifying Message Passing Algorithms Under the Framework of Constrained Bethe Free Energy Minimization},'' \emph{IEEE Transactions on Wireless Communications}, 2021.

\bibitem{pseudoprior}
J.~Goldberger and A.~Leshem, ``{Pseudo Prior Belief Propagation for Densely Connected Discrete Graphs},'' in \emph{IEEE Information Theory Workshop on Information Theory}, 2010.

\end{thebibliography}


\end{document}